\documentclass[reprint, amsmath,amssymb, aps,
]{revtex4-2}

\usepackage[utf8]{inputenc}
\usepackage{ragged2e}
\usepackage{array}
\usepackage{comment}
\usepackage{graphicx}
\usepackage{multirow}
\usepackage{amsmath}
\usepackage{amssymb}
\usepackage{epstopdf}
\usepackage[none]{hyphenat}
\usepackage{xcolor}
\usepackage{hyperref}  
\usepackage{natbib}
\usepackage{tcolorbox}
\usepackage{enumitem}
\usepackage{tikz}
\usepackage{tikz-3dplot}
\usetikzlibrary{positioning}
\hypersetup{colorlinks=true,urlcolor=blue,citecolor=blue,linkcolor=blue}
\usepackage[normalem]{ulem}

\begin{document}
\sloppy
\preprint{APS/123-QED}

\title{Students’ Epistemological Beliefs and their Chatbot Preferences in AI-mediated Physics Learning}

\author{Amogh Sirnoorkar}
\affiliation{Department of Physics and Astronomy, and \\ Department of Curriculum and Instruction, Purdue University, West Lafayette, Indiana - 47907}

\author{Omkar Mamidpalliwar}
\affiliation{Department of Computer Science, \\
Purdue University, Indianapolis - 46202}


\begin{abstract}
Evolving technologies have traditionally influenced pedagogical practices and Generative AI is one such technology that promises to transform higher education. In this study, we investigate the association between introductory students' preferences for chatbot behavior and their epistemological beliefs surrounding physics. Our context involves a custom built online module on waves containing simulations integrated with a chatbot. While students' chatbot preferences were captured through three provided options (guided-inquiry, direct answer, and a combination of inquiry and answer), their epistemological beliefs were captured through the standardized Epistemological Beliefs Assessment for Physical Sciences (EBAPS) survey. Results highlight that students who preferred chatbots that initially engage them in guided-inquiry but provide answers when explicitly sought (`Combination'),  demonstrated sophisticated epistemological beliefs than those who preferred answer-providing chatbots. Notably, we did not observe any association between the EBAPS' total scores among students who preferred guided-inquiry and those who preferred answer-oriented chatbots. Furthermore, the observed differences did not remain statistically significant after applying a Bonferroni-adjusted significance level. Implications of these results for the design and instructional use of chatbots in physics education are discussed.
\clearpage
\end{abstract}

\maketitle

\section{Introduction}\label{sec:intro}

Investigating the processes underlying students’ acquisition of knowledge has been a key focal area in physics education research. From a theoretical standpoint, this domain has largely been examined using ``epistemology'', the study of nature, methods, and justification of knowledge~\cite{elby2001substance,hammer2005resources,chen2022epistemic,sirnoorkar2020towards}.  Epistemological beliefs, which correspond to individuals’ beliefs about knowledge (e.g., its origin or methods of acquisition), represent one of the key components within this broader construct~\cite{hofer1997development,jehng1993schooling,hammer1994epistemological}. For instance, Schommer~\cite{schommer1990effects} proposed five independent beliefs surrounding students’ epistemology: (i) ``certain knowledge'' (absolute knowledge exists and will be eventually uncovered), ``simple knowledge'' (knowledge consists of discrete facts), ``omniscient authority'' (authorities have access to inaccessible knowledge), ``quick learning'' (learning occurs in a quick or not-at-all fashion), and ``innate ability'' (one’s ability to acquire knowledge is endowed at birth).  Epistemological beliefs play a crucial role in students’ learning, as they influence problem-solving~\cite{stathopoulou2007exploring,valanides2005effects}, metacognition~\cite{belet2011meta,bromme2010epistemological}, academic achievement~\cite{aypay2011adaptation,trautwein2007epistemological}, and career choices~\cite{de2007epistemological}.

In this evolving era of generative artificial intelligence (AI), students have access to powerful tools which instantaneously generate and process information, including coursework-related content. Recent AI models have demonstrated the ability to successfully solve typical introductory physics assessments~\cite{kortemeyer2023could,sirnoorkar2024student,bralin2024mapping}, including concept inventories~\cite{west2023ai,polverini2025performance} and Olympiad problems~\cite{qiu2025physics,tschisgale2025evaluating}, while also offering multilingual~\cite{kortemeyer2025multilingual} and multi-modal~\cite{polverini2025multimodal} capabilities. As a result, students are increasingly turning to AI platforms to seek assistance for their coursework~\cite{ravvselj2025higher,fageeh2025rise}. Although this offers convenience, it may also potentially hinder students' learning by reducing their engagement in essential cognitive processes such as modeling, argumentation, and sensemaking~\cite{kosmyna2025your}.  In light of these concerns, researchers have called for investigations into students’ use of AI to better support them in strategically leveraging these tools to enhance, rather than impede their learning.  

Consequently, a growing body of work in PER has sought to unpack students’ practices surrounding their use of AI. These efforts include examining students’ approaches towards seeking feedback from AI~\cite{sirnoorkar2025feedback,mills2025prompting,allen2025students,wan2024exploring}, perceived scientific accuracy and linguistic quality of AI-generated solutions~\cite{dahlkemper2023physics}, impact of AI on students' cognitive load~\cite{lademann2025augmenting}, and pre-service teachers’ use of AI~\cite{hamed2025dual,jiang2026generative,kuchemann2023can}. Additionally, customized chatbots are being developed and deployed for supporting student learning. Hashmi and Rebello~\cite{hashmi2025analyzing} analyzed students' interactions with custom-built chatbot to facilitate Socratic dialogue in the context of physics problem solving. The authors observed students' specificity of questions to increase with every turn of interaction and specificity to correlate positively with self-reported expected course grade.

We find several gaps in the aforementioned literature surrounding students' use of AI, particularly in physics education research. Firstly, research on students' preferences for chatbot behavior, specifically whether students favor chatbots to engage them in guided-inquiry or merely provide answers, remain scarce. Secondly, despite a rich literature on both epistemology and AI in physics education, research exploring the intersection of these two remain relatively underexplored. We contribute to the growing body of literature by exploring the intersection of students' epistemological beliefs and their preferences for chatbot behavior. In doing so, we answer the following research question: {\em What association (if any) exists between introductory students’ preferences for chatbot behavior and their epistemological beliefs in the context of AI-mediated learning of physics?}

This manuscript is structured as follows: in the next section, we provide the relevant background on epistemological beliefs before discussing the study's methodology in Section~\ref{sec:methods}. In Section~\ref{sec:results}, we present the results and discuss them in Section~\ref{sec:discussion}. We conclude the study's implications and limitations in Section~\ref{sec:conclusion}.

\section{Epistemological Beliefs Assessment for Physical Sciences}
\label{subsec:beliefs}

The Epistemological Beliefs Assessment for Physical Sciences (EBAPS) is a 30-item questionnaire consisting of Likert-style and multiple choice questions probing students' epistemological beliefs pertaining to physical sciences~\cite{elby2001helping}~\footnote{https://physics.umd.edu/~elby/EBAPS/home.htm}. Each item is scored on a scale of 0 (least sophisticated belief) to 4 (most sophisticated). The questionnaire probes students’ beliefs across the following five non-orthogonal dimensions (number of items assessing each dimension are indicated in parentheses).

\begin{enumerate}[leftmargin=*]
    \item {\em Structure of scientific knowledge.} (10) Whether scientific knowledge is perceived as a collection of weakly connected pieces of facts, or as a coherent, conceptual, and highly structured unified whole.

    \item {\em Nature of knowing and learning.} (8) The extent to which learning science is viewed as primarily absorbing information, as opposed to actively constructing one’s own understanding by engaging with the material, connecting it to prior experiences and intuitions.

    \item {\em Real-life applicability.} (4) Whether scientific knowledge is perceived as applicable only in formal settings or real-life situations as well.

    \item {\em Evolving knowledge.} (3) The extent to which students balance between the extremes of absolutism (viewing scientific knowledge as fixed and unchanging) and extreme relativism (failing to distinguish between evidence-based reasoning and mere opinion).

    \item {\em Source of ability to learn.} (5) Whether students view success in science as determined by fixed innate ability versus through effort and study strategies. 
\end{enumerate}

\begin{figure}
    \centering    
    \includegraphics[width=0.5\textwidth]{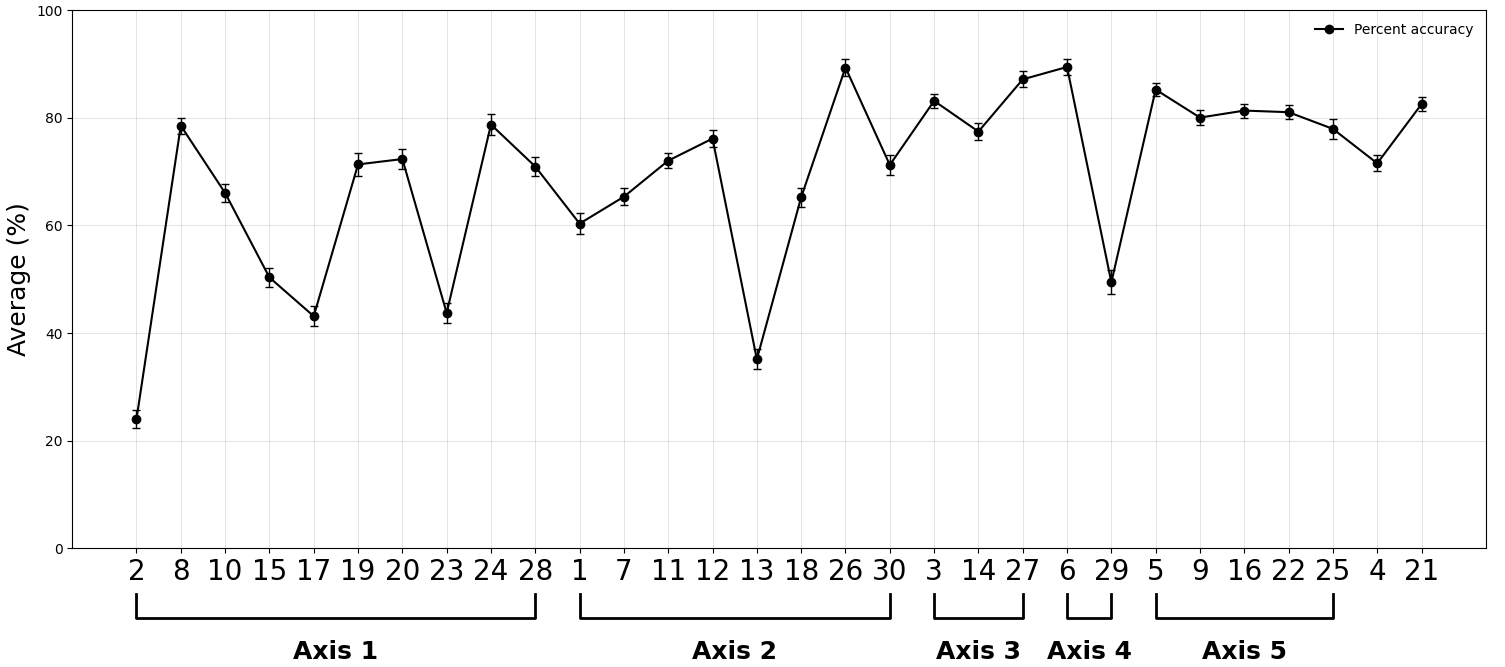}
    \caption{Students’ average percentage scores on EBAPS items, grouped by axis. Items 4 and 21 are not assigned to any of the five axes. Item 19 overlaps between Axes 1 and 3, whereas Item 28 overlaps between Axes 1 and 4.}\label{fig:ebaps-trendline}
\end{figure}

We adopt the EBAPS to capture students’ epistemological beliefs and examine their association with chatbot preferences. The scoring procedure and the resulting measures in our dataset are described below.

\section{Methods}
\label{sec:methods}

Our context involves an extra-credit activity in a a large-enrollment, calculus-based physics course for future engineers in a land-grant midwestern R1 university in the United States. The course follows the Matter and Interactions textbook by Chabay and Sherwood~\cite{chabay2015matter}. The activity focused on the concepts of waves since it was not covered as part of the course and it involved custom-built simulations integrated with a chatbot. The activity had 10 pages. On the first page, students were provided with information about the activity along with instructions to complete it. Students were informed that the activity was was not part of their course assessment and they would receive extra credit for their honest participation. The estimated time for completion was around 30-45 minutes. We captured students' anticipated chatbot preferences in the second page by asking them to highlight one among the following:

\begin{enumerate}
    \item {\bf Direct Answers} (I want the chatbot to give me answers directly rather than guiding me to figure them out on my own.)

    \item {\bf Guided Q\&A} (I want the chatbot to guide me to figure out the questions on my own through conceptual hints and guiding questions.)

    \item {\bf Combination} (I want the chatbot to guide me at first so I can figure out the questions on my own, but provide direct answers if I get stuck for a while.)    
\end{enumerate}

While 52\% highlighted the ``Combination'' option, 39.6\% preferred ``Direct Answers''. Only 7.6\% highlighted the ``Guided Q\&A'' preference. On each of the following pages a specific topic was focused by presenting an interactive simulation and followed by four questions.  On the last page students were asked to provide their student ID, rate the activity on a scale of 5, highlight their perceived learning, and their qualitative feedback about the activity. In addition, we also collected retrospective reflection about their content familiarity and chatbot preferences with the same set of options highlighted above. In the remaining part of this manuscript, we focus only on the information pertaining to the pre-completion of the activity as they highlight students' natural preferences chatbot behavior.  

\renewcommand{\arraystretch}{1.2}
 \begin{table}[tb]
\begin{ruledtabular}
\caption{Details of the students' scores on the EBAPS. }\label{tab:ebaps-stats}

\begin{tabular}{p{0.44\linewidth} p{0.1\linewidth} p{0.16\linewidth} p{0.215\linewidth}}

Axes & Mean & Std. Dev. & Cronbach's $\alpha$ \\
\hline 

1. Structure of knowledge & 2.4 & 1.4 & 0.45 \\

2. Nature of knowing  & 2.7 & 1.3 & 0.42 \\

3. Real-life applicability & 3.2 & 1.1 & 0.33 \\

4. Evolving knowledge & 2.8 & 1.4 & 0.13 \\

5. Source of ability to learn & 3.2 & 1.0 & 0.64 \\

\end{tabular}
\end{ruledtabular}
\end{table}

Before the activity, we also administered the EBAPS survey in the preceding week through Qualtrics~\footnote{https://www.qualtrics.com/} as an extra-credit activity to capture their epistemological beliefs. Both sets of data were collected in Spring of 2026. The total enrollment of the course for the given semester was around 1800. While 1191 responded to the activity, 1408 responded to the EBAPS survey. Across the two sets of data, there were 1048 overlapping responses.

Students' responses  were downloaded and organized into a spreadsheet. The EBAPS responses were scored based on the prescribed rubric. Upon scoring across all 30 items in the survey, students' normalized weighted scores across each of the five  axes were determined in terms of percentages. The weighted score for each EBAPS axis was calculated by dividing the sum of the item scores within that axis by the maximum possible score for that axis. Specifically, if $n_a$ denotes the number of items in axis a and each item is scored on a scale from 0 to 4, the weighted score is computed as:
\begin{equation}
   W_a = \dfrac{\Sigma_{i=1}^{n_a}s_i}{4n_a} \times 100
\end{equation}
where $s_i$ is the score for item $i$. The total score was then determined by summing the weighted scores across the five axes and converting it into percentages. Average scores on each question were then determined to capture the overall trends in students' epistemological beliefs in relation to that of the perfect score (or experts' beliefs). These trends are highlighted in Fig.~\ref{fig:ebaps-trendline}. 

\renewcommand{\arraystretch}{1.2}
 \begin{table}[tb]
\begin{ruledtabular}
\caption{Students'  mean total and axis-wise EBAPS scores (in \%) across the three types of chatbot preferences.}\label{tab:mean-scores}

\begin{tabular}{p{0.1\linewidth} p{0.1\linewidth} p{0.1\linewidth} p{0.1\linewidth} p{0.1\linewidth} p{0.1\linewidth} p{0.1\linewidth}}

Dimension & Inquiry & Answer & Combination \\
      
\hline 

Total & 68.8  & 67.9 & 69.7        \\

Axis-1 & 59.8  & 59.1   & 60.1         \\

Axis-2 & 66.8 & 66.1  & 67.5         \\

Axis-3 & 80.2 & 79.0  & 80.3         \\

Axis-4 & 65.9   & 68.2  & 71.6         \\

Axis-5 &  82.6  & 79.3  &  82.5        \\
\end{tabular}
\end{ruledtabular}
\end{table}

Furthermore, percentage mean and standard deviation values along with Cronbach's alpha~\cite{streiner2003starting} were determined across the five axes to capture the consistency and reliability of students' responses on the survey. These values are tabulated in Table~\ref{tab:ebaps-stats}. The values highlight that, by and large, students demonstrate neither strong (expert-like) nor weak (novice-like) belief structures (based on the mean values) with considerable variance among the responses (based on the standard deviation). The Cronbach's alpha values highlight moderate to low internal consistency among student responses with maximum consistency reported for the fifth axis ({\em Source of ability to learn}) and minimum for the fourth axis ({\em Evolving knowledge}). Though the authors of the survey have argued against this measure for this instrument since ``the assessment items were designed so that students were allowed to disagree with themselves within a subscale'' and because “epistemological beliefs may be triggered depending on context”~\footnote{https://physics.umd.edu/~elby/EBAPS/idea.htm}, we report the values following the tradition in the literature as well as to establish the constraining limits of our results' interpretation based on these scores.

\section{Results}
\label{sec:results}

Table~\ref{tab:mean-scores} provides students' percentage average scores across the total and five axes on the EBAPS survey across the preferences for chatbot behavior. To determine whether any of the three chatbot preference-groups varied across the total and the five axes' scores, we performed the Kruskal-Wallis test (as the scores were not normally distributed) at the nominal significance level of $\alpha = 0.05$. Details of the results are given in Table~\ref{tab:kruskal-results}. The results indicate that at least one pair of chatbot preference groups (Inquiry, Answer, and Combination) differed significantly in their overall EBAPS scores and in their scores on Axis 4 (balance between extremes of absolutism and relativism) and Axis 5 (success in science through innate ability versus effort). No statistically significant differences were observed for Axes 1–3. Furthermore, the small effect sizes indicate that chatbot preference accounted for only a modest proportion of the variability in students' scores. 

Given the observations, we conducted Dunn's post-hoc test (with Bonferroni correction) across the  statistically significant results. For the total score, the test revealed that Answer and Combination groups had statistically different distributions ($p$ = 0.029). Furthermore, students preferring combination-style chatbot had higher EBAPS scores than those who prefer direct answers. No other pairwise differences were statistically significant. 

Among the scores associated with axes 4 and 5, we observed statistical significant difference between the scores associated with Combination and Answer-preferring groups (Axis 4: $p$ = 0.038; Axis 5: $p$ = 0.021). Across both axes, students preferring combination exhibited substantially higher scores than those preferring answer-providing chatbots. No other pairwise differences were statistically significant.

Because six omnibus Kruskal–Wallis tests were conducted (the total EBAPS score and five subscale scores), we additionally evaluated the robustness of these findings using a Bonferroni-adjusted significance level ($\alpha$=0.0083). Under this more conservative criterion, the omnibus differences observed for the total score and Axes 4 and 5 no longer highlighted statistical significance.

\renewcommand{\arraystretch}{1.2}
 \begin{table}[tb]
\begin{ruledtabular}
\caption{Kruskal-Wallis test results across total and axis-wise EBAPS scores among the chatbot preference groups.}\label{tab:kruskal-results}

\begin{tabular}{p{0.2\linewidth} p{0.15\linewidth} p{0.15\linewidth} p{0.2\linewidth} p{0.2\linewidth}}

Dimension & H-stat & $p$-value & Effect size & Significance \\
\hline 
Total & 6.721 & 0.034 & 0.004 & Significant  \\

Axis-1 & 1.293  & 0.523  & 0.000 & Insignificant \\

Axis-2 & 2.761  & 0.251 &  0.001 & Insignificant \\

Axis-3 & 1.842  & 0.398 &  0.000 &  Insignificant  \\

Axis-4 & 9.274  & 0.009 & 0.007  & Significant  \\

Axis-5 & 7.820  & 0.020  & 0.006 &  Significant  \\
\end{tabular}
\end{ruledtabular}
\end{table}

\section{Discussion}
\label{sec:discussion}

In the previous section, we examined the association between students' preferences for chatbot behavior and their epistemological beliefs. At the nominal significance level ($\alpha$=0.05), students who preferred chatbots that initially engaged them in guided inquiry before providing answers exhibited higher EBAPS scores than students who preferred answer-providing chatbots. Differences were observed only for Axes 4 and 5. Specifically, students preferring `Combination' option  exhibited sophisticated beliefs about balancing between absolutism (viewing scientific knowledge as unchanging) and relativism (distinguishing between evidence-based reasoning and mere opinion). These students also demonstrated sophisticated beliefs about the sources of learning ability (effort and effective strategies over innate ability) as compared to those who preferred answer-providing chatbots. However, these differences did not remain statistically significant under the Bonferroni adjustment across the six omnibus comparisons. Thus, the observed patterns are more suggestive than conclusive evidence of an association between chatbot preferences and epistemological beliefs.

Each of these results make unique contributions to the contemporary research on development and deployment of chatbots in STEM education settings. A growing body of contemporary research has focused on exploring the association between students' epistemological beliefs and their interactions with chatbots~\cite{brandle2025comparative,urhahne2026role,sin2025epistemological,avci2025exploring,zhou2025role}. For instance, Br$\ddot{a}$ndle {\em et al.}~\cite{brandle2025comparative} explored the interplay between  epistemological beliefs of ``early adopters'' and ``late users'' of AI  and their AI use. The authors observed that early adopters used AI more frequently and across a wider range of activities, whereas late users exhibited more elaborated beliefs, particularly regarding the source of knowledge. The study did not identify significant correlations between beliefs and AI usage. 

Similarly, Urhahne {\em et al.}~~\cite{urhahne2026role} examined how students' epistemic beliefs about AI (ChatGPT) relate to its adoption in higher education. They found that students who perceived AI as offering diverse ways to justify knowledge claims and who believed it facilitates rapid learning were more likely to use it both instrumentally and as an intelligent learning assistant. Notably, beliefs about knowledge justification most strongly predicted instrumental usage, whereas beliefs about rapid knowledge acquisition most strongly predicted the use of ChatGPT as an intelligent learning assistant. Sin and Joanna~\cite{sin2025epistemological} examined how users' epistemological beliefs predict perceptions and usage of generative AI. They found that stronger beliefs in depending on authority for knowledge were associated with greater familiarity with AI, which in turn indirectly predicted more frequent and more serious AI use. Additionally, stronger beliefs favoring the avoidance of ambiguity were associated with both greater perceived likelihood of AI-related issues and more frequent AI use.

Our study stands apart from the aforementioned studies on several counts. Firstly, while previous studies have primarily examined students' epistemological beliefs about AI or about knowledge more broadly, we investigated their epistemological beliefs specifically pertaining to physics. Secondly, rather than focusing on AI adoption or frequency of use, we examined students' preferences for different chatbot instructional behavior. This distinction shifts our focus from whether students use AI to how they prefer AI to support their learning.

\section{Conclusion}
\label{sec:conclusion}

The evolving capabilities of Generative AI have prompted the need for its integration into STEM learning. This in turn has led to the development of customized chatbots aimed at meeting the goals and expectations of local learning environments. An overarching motivation behind these efforts is to initiate conversations  from a starting point curated to each individual student's choice and elevate it to expected levels of proficiency. In the backdrop of these developments, we explored students' preferences for chatbot behavior and the association of these preferences with regard to their epistemological beliefs.

Our results (though not conclusive and warrant further investigations) provide several implications for the design and instructional use of chatbots in physics education. Firstly, the findings point towards designing chatbots that initially encourage students to reason through a problem but provide more direct assistance when needed. In the context of physics problem solving, custom-built chatbots could begin with questions, hints, or prompts for explanations and then progressively provide more explicit guidance if the student remains stuck. This approach may balance productive struggle with timely support, preventing inquiry from becoming frustrating or answer provision from inhibiting students' engagement in valued epistemic practices such as sensemaking. Secondly, the Axis 5 findings highlight the importance of reinforcing the idea that learning ability can develop through effort, strategy use, feedback, and reflection. When students seek support for physics problems, custom-built chatbots can be designed to recommend a different representation, prompt students to revisit relevant concepts, break a problem into smaller steps, or encourage them to reflect on which strategy was unsuccessful. Instructors can reinforce this message by treating chatbot assistance as support for developing competence rather than as a substitute for thinking. 

Our results also present several implications for further research. We observed beliefs about {\em evolving knowledge} and {\em source of ability to learn} to be potentially associated with students' preferences for chatbots that initially facilitate guided-inquiry but provide answers when required. Further research can examine the potential impact of sustained use of custom-built chatbots facilitating Socratic dialogue on students' epistemological beliefs. In other words, can sustained AI usage initiate positive shifts in students' productive epistemological beliefs or it would have a detrimental impact on students' epistemology? Along the same lines, further research can examine the impact of AI's usage on students' approaches towards scientific inquiry or their engagement in epistemic games while solving physics problems.     

Findings reported in this study accompany several limitations. Firstly, our results are constrained by students' responses on a single survey for beliefs and three narrow choices for chatbot preferences. Detailed qualitative interviews focused on students' beliefs and their chatbot preferences would undoubtedly shine further insights at the intersection of students' epistemology and AI usage. Secondly, the results are also constrained by low values of Cronbach's alpha (especially for Axis 4) and the modest effect sizes. These suggest that the observed associations should be interpreted as preliminary evidence rather than definitive conclusions.

\section{Acknowledgments}
Special thanks to Professor N. Sanjay Rebello, Winter Allen, and Furqan Abbas Hashmi for their insights. This research is supported by National Science Foundation with grant number: 2111138.
\clearpage

\bibliography{ref}
\end{document}